\documentclass[aps,prb,twocolumn,superscriptaddress]{revtex4-1}
\usepackage{graphicx}

\begin{document}
\title{Electronic structure of Ba(Zn$_{0.875}$Mn$_{0.125}$)$_{2}$As$_{2}$ studied by angle-resolved photoemission spectroscopy}

\author{Fengfeng Zhu}
\author{W. X. Jiang}
\author{P. Li}
\author{Z. Q. Yang}
\affiliation{Key Laboratory of Artificial Structures and Quantum Control (Ministry of Education), Department of Physics and Astronomy, Shanghai
Jiao Tong University, Shanghai 200240, China}
\author{H. Y. Man}
\affiliation{Department of Physics, Zhejiang University, Hangzhou 310027, China}
\author{Y. Y. Li}
\author{Canhua Liu}
\author{Dandan Guan}
\author{Jin-Feng Jia}
\affiliation{Key Laboratory of Artificial Structures and Quantum
Control (Ministry of Education), Department of Physics and Astronomy, Shanghai
Jiao Tong University, Shanghai 200240, China}
\affiliation{Collaborative Innovation Center of Advanced Microstructures, Nanjing 210093, China}
\author{F. L. Ning}
\affiliation{Department of Physics, Zhejiang University, Hangzhou 310027, China}
\affiliation{Collaborative Innovation Center of Advanced Microstructures, Nanjing 210093, China}
\author{Weidong Luo}
\email{wdluo@sjtu.edu.cn}
\author{Dong Qian}
\email{dqian@sjtu.edu.cn}
\affiliation{Key Laboratory of Artificial Structures and Quantum
Control (Ministry of Education), Department of Physics and Astronomy, Shanghai
Jiao Tong University, Shanghai 200240, China}
\affiliation{Collaborative Innovation Center of Advanced Microstructures, Nanjing 210093, China}

\date{\today}

\begin{abstract}

Electronic structure of single crystalline Ba(Zn$_{0.875}$Mn$_{0.125}$)$_{2}$As$_{2}$, parent compound of the recently founded high-temperature ferromagnetic semiconductor, was studied by high-resolution photoemission spectroscopy (ARPES). Through systematically photon energy and polarization dependent measurements, the energy bands along the out-of-plane and in-plane directions were experimentally determined. Except the localized states of Mn, the measured band dispersions agree very well with the first-principle calculations of undoped BaZn$_{2}$As$_{2}$. A new feature related to Mn 3d states was identified at the binding energies of about -1.6 eV besides the previously observed feature at about -3.3 eV. We suggest that the hybridization between Mn and As orbitals strongly enhanced the density of states around -1.6 eV. Although our resolution is much better compared with previous soft X-ray photoemission experiments, no clear hybridization gap between Mn 3d states and the valence bands proposed by previous model calculations was detected.
\end{abstract}


\maketitle

Diluted magnetic semiconductors (DMSs) have attracted a lot of attentions due to its great application potential for spintronics devices since the discovery of high-temperature ferromagnetism in Mn-doped GaAs films (Ga$_{1-x}$Mn$_{x}$As, GaMnAs)\cite{Ohno1996a,Ohno1998a,Dietl2000b,Zutic2004a,Dietl2010a,Dietl2014a}. Mn atoms in GaMnAs films have limited chemical solubility and acts as both carrier and spin dopants. The origin of ferromagnetism in GaMnAs remains strongly debated\cite{Dietl2014a}. Recently, a new type of DMSs with decoupled charge and magnetic dopants, Ba$_{1-x}$K$_{x}$(Zn$_{1-y}$Mn$_{y}$)$_{2}$As$_{2}$ (BaKZnMnAs), which is isostructural to "122"-type iron-based superconductors\cite{Paglione2010}, was successfully synthesized with a ferromagnetic (FM) transition temperature (Tc) of $\sim$ 180K\cite{Zhao2013a}. Later on, the highest Tc in BaKZnMnAs was improved to about 230K\cite{Zhao2014c}. The hole carriers are caused by the substitution of Ba by K atoms and the magnetic moments are introduced by Mn atoms in the ZnAs layers. Mn ions were found to be more energetically stable in substitutional sites than in interstitial sites in the ZnAs layers\cite{Glasbrenner2014a}. The independence of the charge and magnetic doping in this system brings a new opportunity to unveil the mechanism of high-temperature ferromagnetic state in DMSs.

Recently, new theoretical model was proposed to understand the BaKZnMnAs system involving the interaction between the localized Mn 3d electrons and the itinerant holes based on the density functional theory (DFT) calculations\cite{Glasbrenner2014a}, in which the low-energy band structure plays a key role. Experimentally, so far, high-resolution band structure measurements were very limited on this system. In previous works, Ba$_{0.7}$K$_{0.3}$(Zn$_{0.85}$Mn$_{0.15}$)$_2$As$_{2}$ samples were studied by soft X-ray absorption and soft X-ray ARPES. Although nondispersive Mn 3d states were observed\cite{Suzuki2015e,Suzuki2015f}, the comparison between the experimental band structure and the first-principle calculations has very large uncertainty\cite{Suzuki2015f} due to limited energy resolution ($\sim$ 150 meV of soft X-ray ARPES). In this work, we carried out vacuum ultraviolet (VUV) ARPES measurements on Ba(Zn$_{0.875}$Mn$_{0.125}$)$_{2}$As$_{2}$ single crystals. High-resolution band dispersions in the ab-plane and along c-axis were determined. We found that the measured band structure agrees very well with DFT calculations, which indicates that correlation effect in this system are very weak. Interestingly, besides the previously reported feature of Mn 3d states\cite{Suzuki2015e}, a new feature of Mn 3d states was observed, which could be closely related to the hybridization between Mn 3d states and As orbitals.

\begin{figure*}[]
\includegraphics[width=17cm]{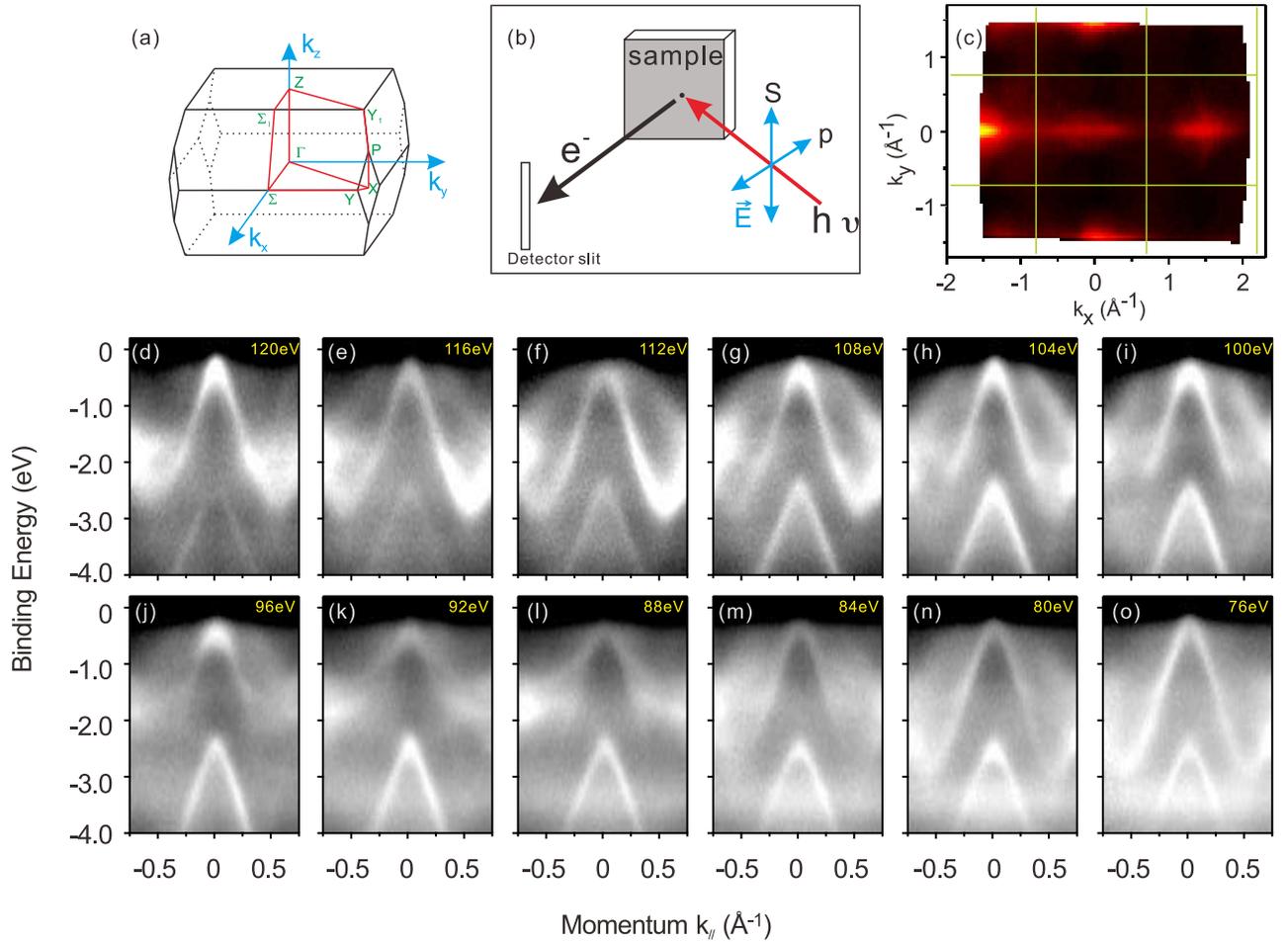}
\caption
{(a) Brillouin zone of $\beta$-BaZn$_2$As$_2$. (b) Sketch of the two light polarizations (S and P-polarization) used in the experiments. (c) Constant energy contour at a binding energy of -0.5 eV. Rectangular symmetry is consistent with the bulk crystal structure. (d)-(o) ARPES spectra along the $\bar{\Gamma}-\bar{\Sigma}-\bar{\Gamma}$ direction using various incident photon energies.
}
\end{figure*}

High-quality Ba(Zn$_{0.875}$Mn$_{0.125}$)$_{2}$As$_{2}$ single crystals were grown by the arc-melting solid-state reaction method. Without K doping, Ba(Zn$_{0.875}$Mn$_{0.125}$)$_{2}$As$_{2}$ is not ferromagnetic\cite{man}.  ARPES experiments were performed at beamline 4.0.3 of Advanced Light Source (ALS) and ARPES beamline of National Synchrotron Radiation Laboratory (NSRL) in Hefei (China). Measurements were carried out at 20 K in ultrahigh vacuum (UHV) with a base pressure better than 1$\times$10$^{-10}$ Torr using Scienta analyzers. The energy resolution is better than 15 meV and angular resolution is better than 1\% of the Brillouin zone (BZ). Photons of 20 ev to 120 eV were used. Figure 1(a) shows the first BZ of BaZn$_2$As$_2$. Two polarizations (S and P) (Fig. 1b) of the incident light were used to change the photoemission matrix element effect\cite{ARPES}. The position of Fermi level was determined by using a polycrystalline Au piece as a reference. Samples were cleaved $\textit{in situ}$ at 20 K to obtain fresh surfaces. Constant energy mapping at binding energy of -0.5 eV (Fig.1(c)) shows a rectangular symmetry, which is consistent with $\beta$-BaZn$_{2}$As$_{2}$ crystal structure\cite{Shein2014a}. The energy bands of BaZn$_2$As$_2$ were calculated using the projector augmented-wave (PAW) method as implemented in the VASP codes\cite{LDA1,LDA2}. Both the cell volume and atomic positions of BaZn$_2$As$_2$ were fully relaxed until the force acting on each ion was less than 0.01 eV/\AA. The optimized lattice constants (a=4.138 \AA \ and c=13.424 \AA) agree well with the experimental values (a=4.131 \AA \ and c=13.481 \AA)\cite{Zhao2013a}. In the calculations, the modified Becke-Johnson (mBJ) exchange potential, same as the previous calculations\cite{Glasbrenner2014a}, was used.

\begin{figure}[]
\includegraphics[width=8cm]{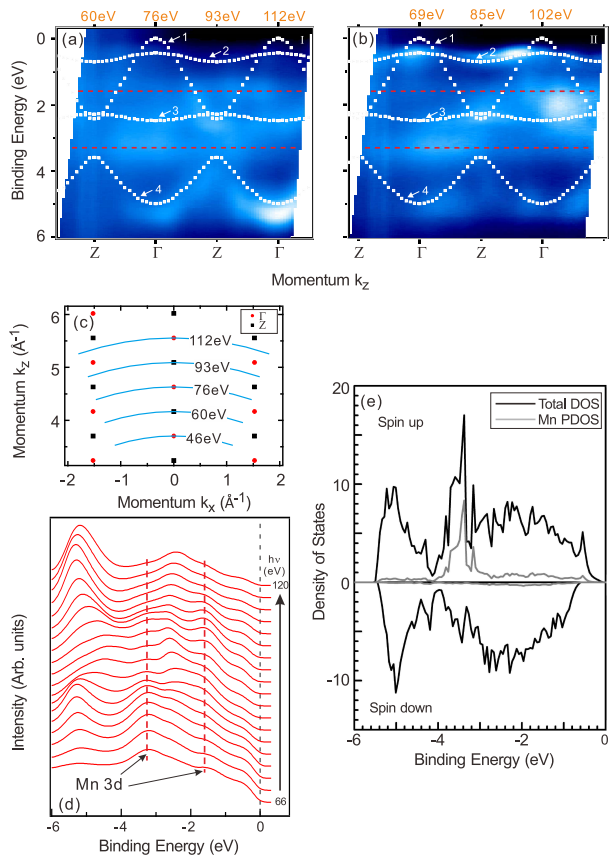}
\caption
{Valence band dispersions along k$_z$ ($\Gamma-Z$) direction measured in the (a) first surface BZ and (b) second surface BZ. White dotted lines are the calculated bands. Red dashed lines indicate two flat bands from Mn 3d states. (c) Locations of $\Gamma$ and Z points in the k$_{x}$ - k$_{z}$ plane. Blue solid lines indicate the momentum positions that can be accessed by changing the incident photo energy and in-plane momentum. (d) some EDCs from (a). Red dashed lines mark the same flat bands in (a). (e) Density of states of Ba(Zn$_{0.875}$Mn$_{0.125}$)$_2$As$_2$ from reference [10]}
\end{figure}

\begin{figure}[]
\includegraphics[width=8cm]{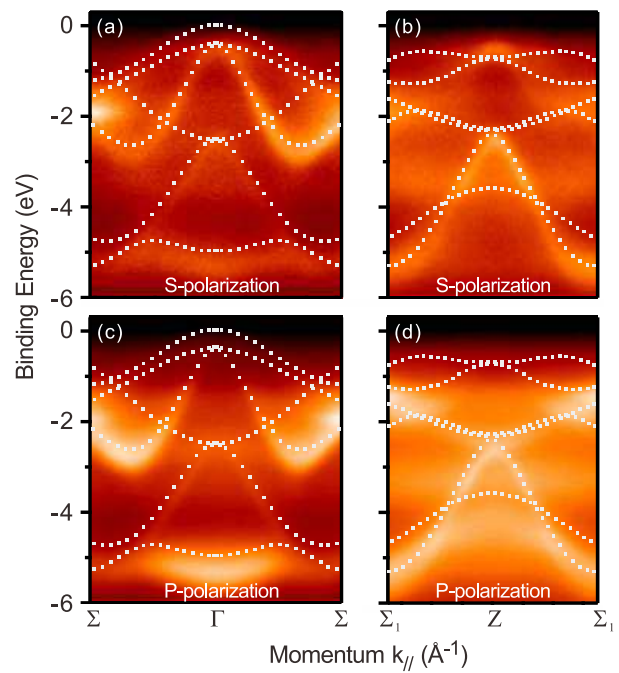}
\caption{The experimental band dispersions in the ab-plane along (a) $\Sigma-\Gamma-\Sigma$ and (b) $\Sigma_{1}-Z-\Sigma_{1}$ directions using S-polarization light. (c) and (d) ARPES spectra along the same directions as (a) and (b) using P-polarization light. White dotted lines represent the calculated band structures of the $\beta$-BaZn$_{2}$As$_{2}$.
}
\end{figure}

One set of ARPES spectra of Ba(Zn$_{0.875}$Mn$_{0.125}$)$_{2}$As$_{2}$ using different incident photon energy ($h\upsilon$ = 120 eV to 76 eV with S-polarization) along the $\bar{\Sigma}-\bar{\Gamma}-\bar{\Sigma}$ direction are shown in Figs. 1(d)-(o). For BaZn$_2$As$_2$, low-energy valence bands mainly come from As 4p orbitals\cite{Glasbrenner2014a} that have low photoemission cross section at VUV energy\cite{Yeh1985}. In previous ARPES study on Ba$_{0.7}$K$_{0.3}$(Zn$_{0.85}$Mn$_{0.15}$)$_2$As$_{2}$ samples, no dispersive features of low-energy valence bands were observed using VUV light\cite{Suzuki2015f}. On our samples, several hole bands centered at $\bar{\Gamma}$ point (k$_{//}$ = 0 \AA$^{-1}$) can be clearly identified, although spectra are extremely weak very close to the Fermi level. Besides the dispersive bands, a non-dispersive feature at binding energy of $\sim$ -3.3 eV was observed (Fig. 1(i)-1(o)). This flat band, also observed in previous soft X-ray ARPES\cite{Suzuki2015e,Suzuki2015f}, belongs to Mn 3d states according to DFT calculations\cite{Glasbrenner2014a}. 

Band dispersion along the k$_z$ direction were experimentally determined using the photon-energy dependent ARPES spectra. Incident photon energies from 56 ev to 120 eV with an interval of 1 eV were used. In order to get accurate results, two sets of data in the 1st and 2nd BZs were collected. We take the energy distribution curves (EDCs) right at k$_{//}$ = 0 \AA$^{-1}$(and k$_{//}$ = 2$\pi$/a in the 2nd BZ) and make new plot based on those EDCs in Fig. 2(a) and 2(b) to show the energy dispersion along the $\Gamma-Z$ direction. Fig. 2(d) shows some EDCs from Fig. 2(a). All EDCs are normalized by the photon flux. More than one BZ along $\Gamma-Z$ direction is covered. The calculated bands of parent BaZn$_{2}$As$_{2}$ are overlaid as the white dotted lines on the experimental spectra. The Fermi level of the calculated bands is slightly adjusted to match the experimental results. Overall the agreement of band energies and dispersions between experiment and calculation is very good except extremely weak intensity very close to Fermi level. In Figs. 2(a) and 2(b), four dispersive bands can be identified. Two bands of large energy dispersion (labelled as band-"1" and band-"4") are more clearly resolvable in the 1st BZ. Another two bands of weak energy dispersion (labelled as band-"2" and band-"4") are visible in the 2nd BZ. Based on the periodicity of the band dispersions observed in experiments, we obtained the inner potential V$_{0}$ $\sim$ 11.5 eV. k$_{z}$ for different photon energy was determined (Fig. 2c). Zone center ($\Gamma$ point in the first BZ) corresponds to $h\upsilon$ = 46, 76 and 112 eV, while zone boundary (Z point in the first BZ) can be accessed by $h\upsilon$ = 93 and 60 eV. Besides these dispersive bands, there are two non-dispersive bands (NB) peaked at $\sim$ -3.3 eV ("NB-1") and -1.6 eV ("NB-2") (marked by the red dashed lines in Figs. 2(a),(b),(d)). NB-1, also observed in soft X-ray photoemission experiments, was assigned to the Mn 3d states\cite{Suzuki2015e,Suzuki2015f}. Using soft X-ray, strong resonance effect on NB-1 was observed near the Mn L-edge. In the VUV region, although the resonance effect is not as strong as L-edge, NB-1 does have enhanced signals around the Mn M-edge (h$\nu \sim$ 60 to 95 eV), shown in Figs. 2(a), 2(b) and 2(d). Previous DFT calculations\cite{Glasbrenner2014a} show that Mn$^{2+}$ state has two main features in the density of states (DOS). Figure 2(e) presents the calculated total DOS and Mn (3d) partial DOS below the Fermi level from reference [10]. In Fig. 2(e), the dominated feature of Mn 3d states is between -3 and -4 eV, which results in the observed NB-1 in ARPES spectra. Beside this main feature, there is another weak feature between $\sim$ -1 to -3 eV in Fig. 2(e). The energy position of the observed NB-2 does match this weak feature in calculations, although the spectra weight is much stronger than calculations. Seen from Fig. 2(a), ARPES intensity of NB-2 are enhanced when h$\nu$ is between $\sim$ 80 and 110 eV. These resonance energies are higher than that of NB-1 ($\sim$ 60 to 90 eV). The different resonant conditions of NB-1 and NB-2 exclude that NB-2 is the surface state of Mn atoms. Therefore, we think that NB-2 should relate to the low energy states of Mn 3d states. In fact, 80 to 110 eV photon energy is between the M-edge of Mn and As, which implies that the observed large spectra weight of NB-2 could be due to the hybridization between Mn and As. Some spectra weight transfers from As to Mn 3d states. Our results suggest that DFT calculations may underestimate the hybridization between Mn and As in the binding energy region of about -1 to -3 eV.

Furthermore, we mapped the band dispersions in the ab-plane along two high-symmetry directions, $\Sigma-\Gamma-\Sigma$ (k$_{z}$ = 0)and $\Sigma_{1}-Z-\Sigma_{1}$ (k$_{z}$ = 2$\pi$/c). Figs. 3(a) and 3(b) show the ARPES spectra measured using incident photo energies of 112 eV and 93 eV with S-polarization, respectively. Fig. 3(c) and 3(d) are the ARPES spectra covering the same momentum space using P-polarization lights. We compared the measured spectra with the calculated bands (white dotted lines) of parent BaZn$_{2}$As$_{2}$. Except the flat features from Mn 3d states, the experimental band structures match very well with the calculations. Shown in Figs. 3(b) and 3(d) where both the valence bands and Mn 3d states have large intensity, we cannot observe any resolvable hybridization gap near the crossing points between the dispersive valence bands and the localized Mn 3d states consistent with X-ray ARPES results\cite{Suzuki2015f}, which implies that the hybridization between itinerant holes and the localized Mn 3d electrons may be not as strong as model calculations expected\cite{Glasbrenner2014a}  at the binding energy of about -3.3 eV.

\begin{figure}[]
\includegraphics[width=8cm]{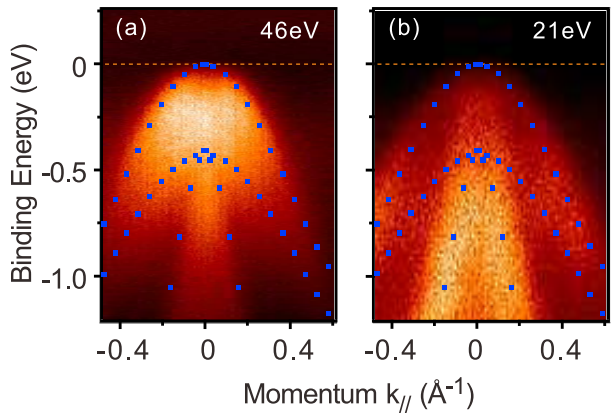}
\caption{
Low-energy band dispersions near the Fermi level. (a) and (b) ARPES spectra taken with h$\nu$ = 46 eV and 21 eV photons with P-polarization. Blue dotted lines are the calculated bands.
}
\end{figure}

In Fig. 3(a), ARPES spectra is extremely weak when the binding energy is very close to Fermi level ($> \sim$ -0.5 eV) due to the photoemission matrix element effect. Fortunately, the low-energy spectra can be revealed to some extent using suitable incident photon energy and polarization. Figure 4 presents the low-energy ARPES spectra in the ab-plane overlaid with the calculated bands (blue dotted lines) at k$_z$ = 4*($4\pi/c$) and 2.91*($4\pi/c$). The measured spectra is consistent with the calculations. Seen from Fig. 4, the Fermi level barely touches the valence band maximum indicating very low carrier density in this nonmagnetic parent compound.

In summary, we have studied the electronic structure of Ba(Zn$_{0.875}$Mn$_{0.125}$)$_{2}$As$_{2}$ by high-resolution ARPES. The measured bands match the DFT calculations of undoped samples very well indicating both weak electron-electron correlation and weak influences of Mn dopants on the primary band structure. Two features related to localized Mn 3d states were observed centered at binding energy of -3.3 and -1.6 eV. The -1.6 eV feature shows larger spectra weight than DFT predication. From the photon energy dependence measurements, we suggests that spectra enhancement around -1.6 eV originates from the hybridization between Mn and As. However, no hybridization energy gap was observed between the itinerate holes and the localized Mn 3d states. Our findings provided important information for exploring the mechanism of the high-temperature FM state in BaZnMnAs system.

This work is supported by National Basic Research Program of China (Grants No. 2012CB927401, No. 2013CB921902), National Natural Science Foundation of China (Grants No. 11574201, No. 11521404, No. 11134008, No. 11174199, No. 11374206, No. 11274228, No. 11227404, No. 91421312, and No. 91221302), Shanghai Committee of Science and Technology (No. 12JC140530). D.Q. acknowledges support from the Changjiang Scholars Program and the Program for Professor of Special Appointment (Eastern Scholar). The Advanced Light Source is supported by the Director, Office of Science, Office of Basic Energy Sciences, of the US Department of Energy under Contract No. DE-AC02-05CH11231.

\end{document}